\documentstyle[12pt,epsf]{article}

\def\be{\begin{equation}}
\def\ee{\end{equation}}
\def\bea{\begin{eqnarray}}
\def\eea{\end{eqnarray}}

\newcommand{\Section}[1]{\section{#1}\setcounter{equation}{0}}

\def\nn{\nonumber}

\begin{document}

\begin{flushright}
IPM/P-98/24  \\
cond-mat/9812054
\end{flushright}

\pagestyle{plain}
\vskip .05in                                   
\begin{center}

\Large{\bf Riemannian geometry of the Pauli paramagnetic gas}
\small
\vskip .15in

Kamran Kaviani ${}^{a,b}$, Ali Dalafi Rezaie ${}^c$\\
\vspace{.5 cm}
{\it a)Department of Physics, Az-zahra University,\\
P.O.Box 19834, Tehran, Iran}\\
\small
{\it b)Institute for Studies in Theoretical Physics and Mathematics (IPM),}\\
{\it P.O.Box 19395-5531, Tehran, Iran}\\
\vspace{.3 cm}
{\it c)Department of Physics, Tehran University,\\
P.O.Box 14394, Tehran, Iran}\\
\vspace{.3 cm}
{\sl E-mails:  kaviani@theory.ipm.ac.ir}

\begin{abstract}
  We investigate the thermodynamic curvature resulting from a Riemannian 
geometry approach to thermodynamics for the Pauli paramagnetic gas which is
a system of identical fermions each with spin ${1\over 2}$.
We observe that the absolute value of thermodynamic curvature can 
be interpreted as a measure of the  
stability of the considered system only in the classical and 
semiclassical regime.
But in quantum regime some exceptions are observed.
\end{abstract}
\end{center}

\Section{Introduction}
 Thermodynamic fluctuation theory whose basic goal is to express the time
independent probability distribution for the state of a fluctuating system
in terms of thermodynamic quantities, is usually attributed to Einstein who
applied it to the problem of blackbody radiation [1].
 The full formalism for classical thermodynamic fluctuation theory was worked 
out by Green and Callen [2] in 1951 and elaborated upon by Callen [3].

However, despite of a wide range of applicability, the classical fluctuation
theory fails near critical points and at volumes of the order of the 
correlation volume and less.

 In 1979 Ruppeiner [4] introduced a Riemannian metric structure representing
thermodynamic fluctuation theory, and related to the second derivatives of
the entropy. His theory offered a good meaning for the distance between 
thermodynamic states. He showed that the breakdown of the classical theory
occurs because it does not take into account local correlations[5]. This 
deficiency of the classical theory is precluded in the covariant fluctuation
theory of Ruppeiner by using a hierarchy of concentric subsystems, each of
which samples only the thermodynamic state of the subsystem immediately
larger than it[6,7].
 One of the most significant topics of this theory is the introduction of
the Riemannian thermodynamic curvature as a qualitatively new tool for the
study of fluctuation phenomena. It is this geometry which is the basis for
this paper. Here we investigate the case of a Pauli paramagnetic gas that
is a gas of identical spin $1\over 2$ fermions in the presence 
of an external magnetic field.

 The outline of this paper is as follows. First the Riemannian geometry of
thermodynamic fluctuation theory is summarized. Second, the Riemannian scalar
curvature of the Pauli paramagnetic gas is evaluated. Finally the curvature
of the classical ideal paramagnetic gas is calculated and is compared to the
the curvature of Pauli paramagnetic gas in the classical limit.
\Section{Geometrical view point of thermodynamics}
In this section we review the Riemannian geometry of thermodynamics,
discuss its connection to the covariant thermodynamic fluctuation theory,
and summarize the resulting interpretation of thermodynamic curvature.
Riemannian structure of the thermodynamic state space is defined by the
second derivatives of a thermodynamic potential density as a metric tensor[7].
If we choose extensive densities as coordinates, we can use either energy
or entropy density as the potential and these two descriptions are 
thermodynamically equivalent; but the metric tensor is different in these two
representations.
Here we work in entropy representation, because then the meaning of the
distance, measured in units of average fluctuations, is very transparent.
When some extensives are substituted by intensives, the potential is a
Massieu function[3].
   Now we consider an open subsystem $A_V$, with fixed volume $V$, of a
       thermodynamic
fluid system $A_{V_0}$ with a very large volume $V_0$. The system $A_{V_0}$
consists
of $r$ fluid components and is in equilibrium. We denote by the $n$-tuple
$a_0 = (a^0_0,a^1_0,a^2_0,\cdots , a^r_0)$
the internal energy per volume and the number of particles per volume of the
$r$ components of $A_{v_0}$[7]. These parameters are the standard densities 
in the
entropy representation; they constitute the thermodynamic state of $A_{V_0}$. The
subsystem $A_V$ has the corresponding thermodynamic state $a$.
   The Gaussian approximation of the classical thermodynamic fluctuation
theory asserts that the probability of finding the thermodynamic state of
$A_V$ between $a$ and $a+da$ is[7]:
\bea
P_V(a|a_0) da^0 da^1 \cdots da^r &=& 
\left({V \over {2 \pi}}\right)^{{r+1}\over 2}
\exp [-{V\over 2} g_{\mu\nu}(a_0) \Delta a^\mu \Delta a^\nu ] \nn \\
&\times & \sqrt{g(a_0)} da^0 da^1 \cdots da^r,
\eea
where
\bea
\Delta a^\mu &=& a^\mu - a_0^\mu \nn \\
g_{\mu\nu}&=& \left.- {1\over K_B} {\partial^2s\over {\partial a^\mu
\partial a^\nu}}\right|_{a=a_0},
\eea
where $s$ is the entropy per volume in the thermodynamic limit, $K_B$ is
Boltzman's constant, and $g(a_0)= \det[g_{\mu\nu}(a_0)]$.

The quadratic form in Eq.(2.1),
\be
(\Delta l)^2 = g_{\mu\nu}(a_0) \Delta a^\mu \Delta a^\nu
\ee
constitutes a positive definite Riemannian metric on the thermodynamic state
space. The positive definiteness results since the entropy is a maximum in
equilibrium $a=a_0$. Eqs.(2.1) and (2.3) denote the Physical interpretation
for the distance
between two thermodynamic state. The less the probability of a fluctuation
between the states, the further apart they are. The quantity
$$
\sqrt{g(a_0)} da^0 da^1\cdots da^r
$$
in Eq.(2.1),
is the invariant Riemannian thermodynamic state space volume element.
The form of Eq.(2.2) holds only in standard densities. To express the metric
tensor in a general set of thermodynamic coordinates $x=x(a)$, 
one can use the following transformation rule
\be
g'_{\alpha \beta}(x)={{\partial a^\mu} \over{\partial x^\alpha}}
{{\partial a^\nu} \over{\partial x^\beta}}g_{\mu\nu}(a).
\ee
Having the metric we can calculate the Riemannian curvature tensor. For our
metric, the scalar curvature $R$ ,has units of real space volume, regardless
of the dimension of the state space[11]. 
It is a measure of effective interaction
between the components of the system, proportional to the correlation volume,
and diverges near the critical point of the pure interacting fluid.
   Covariant thermodynamic fluctuation theory indicates that curvature is a
measure of the smallest volume where classical thermodynamic fluctuation
theory could work. This theory was proposed as the correct way to extent the
classical thermodynamic fluctuation theory beyond the Gaussian 
approximation[7].

An alternative interpretation of the thermodynamic curvature was offered by
Janyszek and Mrugala[8]. They suggested that the thermodynamic curvature
is a measure of the stability of the considered system. The system is less
stable if the curvature increases and vice versa. 
Also these authors calculated
the curvature of ideal Fermi and Bose gases[9]. They show that these systems
have the curvature with opposite signs.

In this paper we interpret the absolute value of the curvature as a measure
of stability in order to come to an agreement with the curvature of the boson
ideal gases that diverge to negative infinity (in the sign convention used
here) where the Bose-Einstein condensation occurs[9].

\Section{Geometry of the Pauli paramagnetic gas}
   We now turn our attention to studying the equilibrium state of a gas of
noninteracting fermions in the presence of an external magnetic field $H$.

   The extensive parameter which describes the magnetic properties of a
    system
is $M$, that is the component of the total magnetic moment
parallel to the external field.
   The entropic intensive parameters are defined as[3],
\be
F^1={\partial S \over \partial U}= {1\over T}; \;\;\;\;\;\;\;\;\;\;
F^2={\partial S \over \partial N}= -{\mu\over T}; \;\;\;\;\;\;\;\;\;\;
F^3={\partial S \over \partial M}= -{H\over T}.
\ee

   We use the thermodynamic potential $\phi$ which is defined as,
\be
\phi = s [{1\over T}, -{\mu \over T}, -{H\over T}]=s-
{1\over T}u +{\mu \over T}\rho + {H\over T}m = {P\over T},
\ee
where $u$,$\rho$,$m$ and $P$ are energy per volume, density,
magnetization and pressure respectively.
The energy of a particle, in presence of an external magnetic field $H$,
is given by
\be
{\cal E}={p^2\over {2 m_0}} - \vec J. \vec H
\ee
where $\vec J$ is the intrinsic magnetic moment of the particle and $m_0$
is its mass.
   For the case of the Pauli paramagnetic gas the spin of each particle
    is $1\over 2$; the vector
    $\vec J$
must then be either parallel to the vector $\vec H$ or anti parallel.
 From the grand canonical distribution (using Fermi-Dirac statistics)
  one can obtain the following
  equations[13]:
\bea
\ln Q &=& {{P V}\over {K_B T}} = {V\over \lambda^3}(f^+_{5\over 2}
+f^-_{5\over 2}), \\
\rho &=& {N \over V} = {1\over \lambda^3} (f^+_{3\over 2} + f^-_{3\over 2})
\eea
where 
\bea
f^\pm_n &=& f_n(\eta^\pm);  \\ 
\eta^{\pm} &=& \eta \exp [\mp {{ J H} \over {K_B T}}]
=\exp [{\mu \over {K_B T}}\mp {{ J H} \over {K_B T}}]
\eea
and $\lambda = {h \over (2 \pi m_0 K_B T)^{1\over 2}}$ is the mean 
thermal wavelength of the particle, $h$ is the Planck
constant and
\be
f_n(\eta ) = {1\over \Gamma (n)} \int_0^\infty {{X^{n-1} dX}\over {e^X\over 
\eta } +1}
\ee
we have used the standard symbol for the fugacity
$\eta = \exp ({\mu \over {K_B T}})$.
From Eqs.(3.2) and (3.4) the thermodynamic potential is obtained
\be
\phi (x,y,z) = I x^{-{3\over 2}}[f_{5\over 2}(e^{-y-Jz})+ f_{5\over 2}(
e^{-y+Jz})].
\ee
Where $I={{(2\pi m)^{3\over 2}}\over h^3}$ and
$x=F^1, y = F^2, z= F^3 $.
We have set $K_B=1$.

   Now it is straightforward to obtain the metric elements in $F$
    coordinates[7]:
\be
g_{\mu\nu}={\partial^2 \phi \over {\partial F^\mu \partial F^\nu}}
\ee
   According to Eqs.(3.9) and (3.10) and noting that,
${\partial  f_n(\eta ) \over {\partial \eta }} 
={1 \over \eta} f_{n-1}(\eta )$,
the components of the metric tensor are as follows
\bea
g_{11}&=&{15\over 4} I x^{-{7\over 2}} (a + b)\;\;\;\;\;\;\;\;\;\;
g_{12}={3\over 2} I x^{-{5\over 2}} (c + d)\;\;\;\;\;\;\;\;\;\;
g_{13}=-{3\over 2} I J x^{-{5\over 2}} (c - d) \nn \\ 
g_{22}&=& I x^{-{3\over 2}} (e + f)\;\;\;\;\;\;\;\;\;\;\;\;\;\;
g_{23}= - I J x^{-{3\over 2}} (e - f)\;\;\;\;\;\;\;
g_{33}= I J^2 x^{-{3\over 2}} (e + f)
\eea
where $a= f^+_{5\over 2}, \; b= f^-_{5\over 2}, \; c=f^+_{3\over 2}, \;
d= f^-_{3\over 2}, \; e=f^+_{1\over 2}, \; f=f^-_{1\over 2}$;
they are functions of $y$ and $z$. Their derivatives with respect to $y$ and
$z$ are as follows:
\be
\matrix {
{\partial a \over \partial y} = -c & {\partial a \over \partial z} =Jc &
{\partial b \over \partial y} = -d & {\partial b \over \partial z} =-Jd \cr
{\partial c \over \partial y} = -e & {\partial c \over \partial z} =Je &
{\partial d \over \partial y} = -f & {\partial d \over \partial z} =-Jf \cr
{\partial e \over \partial y} = -h & {\partial e \over \partial z} =Jh &
{\partial f \over \partial y} = -k & {\partial f \over \partial z} =-Jk \cr}
\ee
where $h= f^+_{-{1\over 2}}, \; k=f^-_{-{1\over 2}}$, 
these quantities are used for obtaining the scalar
curvature.
   The Riemann and the Ricci tensors and the scalar curvature
   are respectively
\bea
R^\kappa_{\lambda\mu\nu}&=& \partial_\mu \Gamma^\kappa_{\nu\lambda}-
\partial_\nu
\Gamma^\kappa_{\mu\lambda} + \Gamma^\eta_{\nu\lambda} \Gamma^\kappa_{\mu\eta}
- \Gamma^\eta_{\mu\lambda} \Gamma^\kappa_{\nu\eta} \nn \\
Ric_{\mu\nu}&=& R^\lambda_{\mu\lambda\nu}\nn\\
R&=&g^{\mu\nu}Ric_{\mu\nu}
\eea
where $\Gamma$s  are the Christoffel symbols[14].
   The scalar curvature may be worked out with Eqs.(3.11),(3.12),(3.13):
 \bea
 R&=&  {\lambda^3\over {2 (5 e f a +5 e f b - 3 c^2 f -
 3 d^2 e )^2}}
 [55 f^2 a e^2 + 55 f^2 b e^2 \nn \\
  &-& 28 f^2 e c^2 - 25 f^2 a c h  - 25 f^2 b c h - 28 f d^2 e^2
 + 12 f d^2 c h \nn \\
  &-& 25 e^2 d k a - 25e^2 b d k
 + 12 c^2 d k e + 15 c d h k a   + 15 c d b h k ]
 \eea
As it is seen from Eq.(3.14), $R$ is a symmetric function of $z$; this means
that scalar curvature is independent of the orientation of external magnetic
field, $R(-H)=R(H)$.

In the classical limit in the absence of the external magnetic field, we have
$\eta^\pm \rightarrow \eta \rightarrow 0$ and $f^{\pm}_n(\eta) \rightarrow
\eta$; so $R$ is obtained as follows:
\be
R= {1\over 4} {\lambda^3\over\eta}
\ee
On the other hand in this limit, Eq.(3.5) results
\be
\rho = {2\over{\lambda^3}} \eta 
\ee
From Eqs.(3.15) and (3.16) the classical limit of $R$ is given by
\be
R= {1\over{2 \rho}}
\ee
This surprising simple result, shows that in the classical limit the scalar
curvature is in the order of the volume occupied by a single particle.
It is
in complete agreement with the scalar curvature obtained by Ruppeiner for
multicomponent ideal gas[11]. It means that in the classical limit the
scalar curvature of the Pauli paramagnetic gas behaves like that of a
two-component ideal gas.

In Fig.1  we present the dependence of $R$ on $\eta$ for a fixed value
of $H$, for an isotherm in units of $\lambda^3$.
In the classical region where $\eta<1$, $R$ diverges near $\eta=0$.
It relates to the fact that in this limit $\rho$ goes to zero (as Eq.(3.17)
demonstrates), so there are not enough particles for a continuous
thermodynamic description. In the quantum mechanical region, where
$\eta \gg 1$, $R$ tends to a constant value.

Fig.2 shows the dependence of $R$ on $H$ for a fixed value of $\eta$,
for an isotherm in units of $\lambda^3$. $R$ is a monotonically decreasing
function of $H$. Physically, as the external magnetic field increases,
the magnetization fluctuation decreases. So the system becomes more stable.
Here we can interpret $R$ as a measure
of the stability of the thermodynamic system.
The less the magnitude of $R$, the more stable the system becomes.

\begin{center}
\begin{figure}
\centerline{
\epsfxsize=7cm
\epsfbox{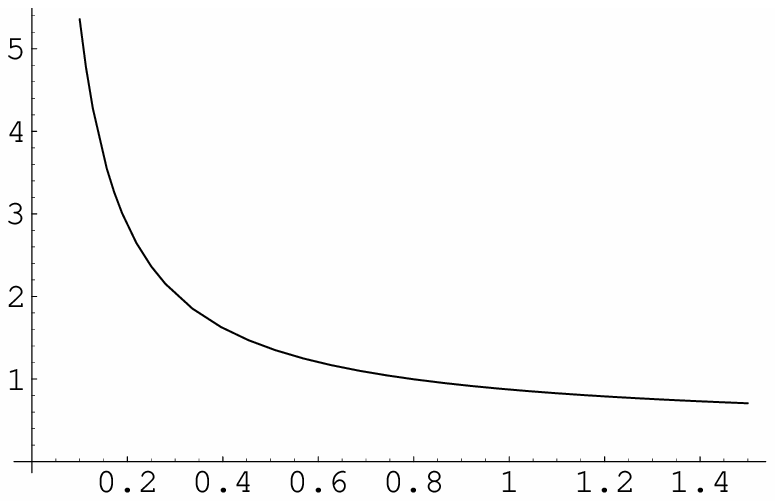}
\epsfxsize=7cm
\epsfbox{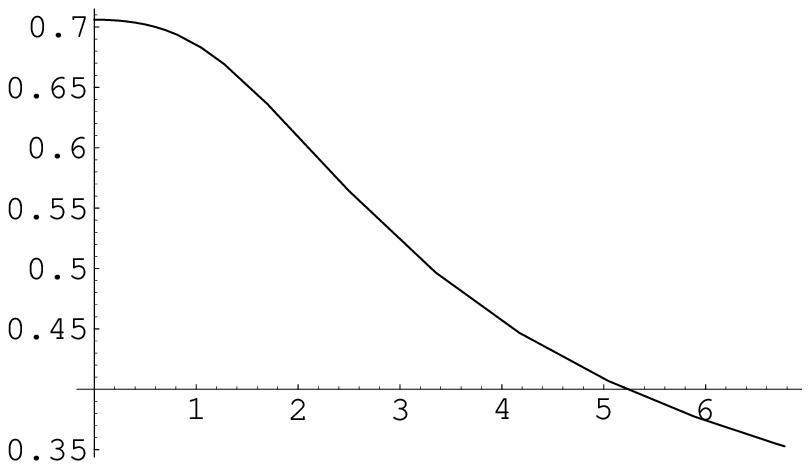}}
\hskip 3cm
Fig.1
\hskip 6cm
Fig.2
\\
{\vskip -4cm
\hskip 3cm
$z=1$
\hskip 7cm
$\eta=1.5$}
\vskip 1cm
\hskip 6.4cm
$\eta$
\vskip -1cm
\hskip 13.7cm
$z$
\vskip -3.5cm
\hskip .5cm
${R\over{\lambda^3}}$
\hskip 7cm
${R\over{\lambda^3}}$
\end{figure}
\end{center}
\begin{figure}
\vskip 5cm
\hskip 1cm In Fig.3 there is something new that we can not explain 
it in terms of
fluctuations. This figure shows the dependence of $R$ on $H$ for $\eta = 10$.
As it is seen in this quantum regime, $R$ has maximum around $z=2$. Fig.4
shows that this maximum of $R$ occurs at higher values of $z$ as we increase
the value of $\eta$. This behavior is in contrast with our intuition; because
the magnetization fluctuation, ${<(\Delta m)^2>}\over {<m>^2}$ is a
monotonically decreasing function of $H$ for all values of $\eta$. So it
seems that the stability interpretation of thermodynamic curvature fails in
strong quantum regime.
\vskip 1cm
\centerline{
\epsfxsize=7cm
\epsfbox{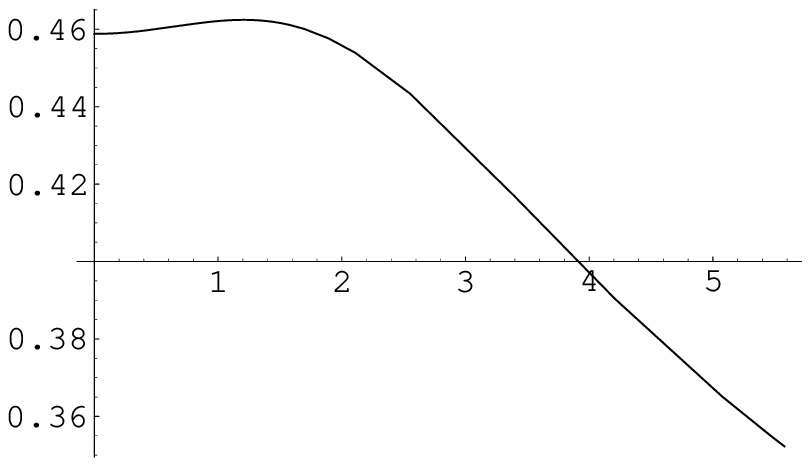}
\epsfxsize=7cm
\epsfbox{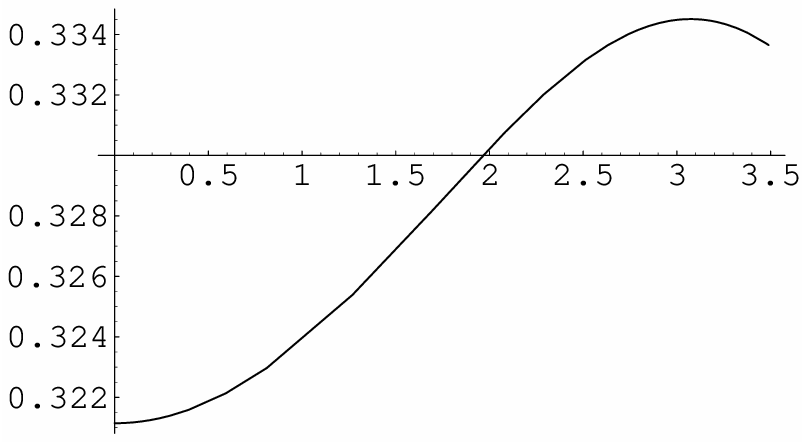}}
\hskip 3cm
Fig.3
\hskip 6cm
Fig.4
\\
{\vskip -5cm
\hskip 4cm
$\eta=10$
\hskip 7cm
$\eta=40$}
\vskip .65cm
\hskip 6.5cm
$z$
\vskip -1.15cm
\hskip 13.7cm
$z$
\vskip -2cm
\hskip .5cm
${R\over{\lambda^3}}$
\hskip 7cm
${R\over{\lambda^3}}$
\end{figure}
\vskip 4cm
For the last point we allude to a relationship between 
$R$ and the correlation
volume. We note that the correlation function of Fermi gas 
in classical regime
$(\rho \lambda^3 \ll 1\; {\rm or }\;\eta \ll 1)$ is given by the formula[12]
\be
\nu(r)=-{1\over 2} e^{-{2\pi r^2}\over \lambda^2}.
\ee
So one can see that the correlation volume in this regime is
\be
V_{cor}={\lambda^3 \over {(2\pi)^{3\over 2}}}.
\ee
It means that in the classical regime $R\eta$ is proportional to the
correlation volume. The possible relationship between the curvature and the
correlation volume in the quantum regime has not been explored.

\Section{Classical ideal paramagnetic gas}
In this section we calculate the scalar curvature of a classical paramagnetic
gas
and compare it to the scalar curvature of the Pauli paramagnetic gas in the
limit of low fugacity and low magnetic fields.

Consider a gas of identical, mutually noninteracting and freely orientable 
dipoles, each having a magnetic moment $J$. In the presence of an external 
magnetic field $H$, the dipoles experience a torque tending to align them in 
the direction of the field. The energy of a particle is given by
\be
{\cal E}={p^2 \over{2m_0}} - J H \cos{\theta}
\ee
Here we have neglected the effect of the induced magnetic field.
Mijatovic et al used the {\it energy form} of the metric to evaluate the
geometry in the paramagnetic ideal gas[10]. 
They also used the particle number
as the fixed scale. Here we use the {\it energy form} of the metric. we also
use the volume as fixed scale.
The logarithm of the grand canonical partition function
(using Maxwell-Boltzman statistics) is obtained as
follows:
\be
\ln Q_c ={{P V}\over {K_B T}}= 4 \pi \eta {V\over{\lambda^3}} {{\sinh (Jz)}
\over{Jz}}
\ee
and the thermodynamic potential is again obtained from Eqs.(4.2) and (3.2)
\be
\phi_c = 4\pi I x^{-{3\over 2}} e^{-y} {{\sinh (Jz)}\over{Jz}}
\ee
Where x,y and z are those that are defined in section 3 and $K_B=1$. 
From Eq.(3.10) the metric
elements are obtained
\bea
g_{11} &=& 15 \pi I x^{-{7\over 2}}e^{-y} {{\sinh (J z)}\over{J z}}  \cr
g_{12} &=& 6 \pi I x^{-{5\over 2}}e^{-y} {{\sinh (J z)}\over{J z}} \cr
g_{13} &=& -6 \pi I x^{-{5\over 2}}e^{-y} ({{\cosh (J z)}\over {z}} - {{\sinh (J z)}\over{J z^2}}) \cr
g_{22} &=& 4 \pi I x^{-{3\over 2}}e^{-y} {{\sinh (J z)}\over{J z}} \cr
g_{23} &=& -4 \pi I x^{-{3\over 2}}e^{-y} ({{\cosh (J z)}\over {z}} - {{\sinh (J z)}\over{J z^2}})\cr
g_{33} &=& 4 \pi I x^{-{3\over 2}}e^{-y} ({{J \sinh (J z)}\over z}
- 2 {{\cosh (J z)}\over z^2}\cr &+& 2 {{\sinh (J z) }\over{ J z^3}})
\eea

Using Eqs.(3.13) and (4.4), one can calculate the scalar curvature\\
\be
R_c = {1\over {8 \pi}}{\lambda^3\over \eta} {{J z}\over {\sinh J z}}
\ee
Eqs. (4.3) and (4.5) show clearly that $R_c$ and $\phi_c$ 
satisfy the following equation.
\be
R_c=\kappa {K_B\over \phi_c}
\ee
where $\kappa ={1\over 2}$ and $K_B=1$.
This interesting result is nothing except the geometrical equation with 
$\kappa ={1 \over 2}$[7].

On the other hand , the equation of state of the classical ideal gas
$(P V = N K_B T)$ and Eqs.(4.6) and (3.2) results
\be
R_c ={1\over {2\rho}}
\ee
Eq.(4.7) shows that the curvature of the classical ideal paramagnetic gas is
in order of the volume occupied by a single particle.

We can see the magnetic field dependence of $R_c$ through Eq.(4.5). It is
a monotonically decreasing function of $z$ and has a maximum at $z=0$.
Here we can interpret the curvature as a measure of stability, since the
magnetization fluctuation, ${{<(\Delta m)^2>}\over{<m>^2}}$, is dominant
near $z=0$, and decreases monotonically as $z$ increases.

Now let us look at the Eq.(3.14) for the curvature of the Pauli gas in the
limit of low fugacity and low magnetic field, where
$f^{\pm}_n \rightarrow \eta^{\pm}$ then one can obtain
\be
R= {1\over 4} {\lambda^3 \over \eta} {1\over{\cosh J z}}.
\ee
Eq.(4.8) is similar to Eq.(4.5), because the behavior of $\cosh z$
is similar to
that of ${\sinh z} \over z$; but they are somewhat different. The source of 
this difference is related to the fact that in the case of the classical 
ideal paramagnetic 
gas, each dipole is freely orientable whereas each particle in the Pauli 
paramagnetic gas can choose only two directions (even in the limit of low
fugacity).\\

In fact, had we constrained the dipoles to choose only two directions 
(parallel or anti parallel to H ) i.e. set $\cos \theta=\pm1$ in Eq.(4.1) and
used the classical grand partition function we would have obtained
\be
\ln Q'_c =2\eta {V\over \lambda^3} \cosh J z.
\ee

This is just Eq.(3.4) in the limit of low fugacity and low magnetic 
fields (where $f^\pm_n \rightarrow \eta^{\pm}= \eta e^{\pm zJ}$).
The thermodynamic potential is
\be
\phi'_c = 2 I x^{-3\over 2} e^{-y} \cosh J z.
\ee
Again this is the classical limit of Eq.(3.9). Using Eqs.(3.10) and 
(3.13) one can obtain the scalar curvature which is just Eq.(4.8)
\Section{Conclusion}
We have evaluated the thermodynamic curvature for the Pauli paramagnetic gas
that is a system consisting of identical  spin ${1\over 2}$ fermions.

In the classical limit (i.e $\eta \ll 1$) in the absence of the external
magnetic field this curvature reduces to that of a two-component ideal gas.
In this regime we can find a simple relationship between the curvature and
the correlation volume.

In the limit of low fugacity and for a finite value of external magnetic 
field
the curvature of the Pauli gas coincides with that of the classical ideal
paramagnetic gas which is a monotonically decreasing function of magnetic
field; here we can interpret the curvature as a measure of stability.
The system becomes more stable if the absolute value of the curvature 
decreases.

In the quantum mechanical regime (where $\eta \gg 1$) the curvature as a
function of magnetic field has a maximum. This maximum occurs at stronger
magnetic fields as the magnitude of fugacity is increased. Stability
interpretation of curvature can not explain this maximum.
\\
\\
\section *{Acknowledgment}
We would like to thank M. Khorami for useful discussions and comments.

\end{document}